\documentclass[useAMS,usenatbib]{mn2e}
\usepackage{graphicx}
\usepackage{longtable}
\usepackage{lscape}
\usepackage{overpic}
\usepackage[T1]{fontenc}
\usepackage{aecompl}
\usepackage[normalem]{ulem}

\newcommand\nion[2]{#1\,\lowercase{{\sc #2}}}

\def\kmsec{${\rm km~s^ {-1}}$}
\def\BmV0{\mbox{$(B-V)^{\rm 0}$}}
\def\VmK0{\mbox{$(V-K)^{\rm 0}$}}
\def\MV0{\mbox{$M_{\rm V}^{\rm 0}$}}

\def\simgt{\lower.5ex\hbox{$\; \buildrel > \over \sim \;$}}
\def\simlt{\lower.5ex\hbox{$\; \buildrel < \over \sim \;$}}

\def\feh{\rm[Fe/H]}
\def\mgfe{\rm[Mg/Fe]}

\def\afe{\rm[\alpha/Fe]}

\def\ezec{E_{\rm z}/E_{\rm c}}
\def\jzjc{J_{\rm z}/J_{\rm c}}
\def\MA{$\left[[{\rm Fe/H}],[{\rm Mg/Fe}]\right]$}
\def\EzJz{$[J_z/J_c,E_z/E_c]$}

\title{The hunt for the Milky Way's accreted disc}

\author[Ruchti et al.]{Gregory~R.~Ruchti,$^{1}$\thanks{{\tt email: greg@astro.lu.se}} Justin~I.~Read,$^{2}$ Sofia~Feltzing,$^{1}$ Antonio~Pipino$^{3}$
\newauthor
and Thomas~Bensby$^{1}$ \\
$^{1}$Lund Observatory, Department of Astronomy and Theoretical Physics, Box 43, SE-22100, Lund, Sweden\\
$^{2}$Department of Physics, University of Surrey, Guildford, GU2 7XH, Surrey, UK\\
$^{3}$Institute for Astronomy, Department of Physics, ETH Z\"urich, Wolfgang-Pauli-Strasse 16, 
CH-8093, Z\"urich, Switzerland
}

\date{Accepted 2014 July 15.  Received 2014 July 15; in original form 2014 March 11}

\begin{document}

\maketitle

\begin{abstract}
The Milky Way is expected to host an accreted disc of stars and dark matter. This forms as massive $\simgt 1:10$ mergers are preferentially dragged towards the disc plane by dynamical friction and then tidally shredded. The accreted disc likely contributes only a tiny fraction of the Milky Way's thin and thick stellar disc. However, it is interesting because: (i) its associated `dark disc' has important implications for experiments hoping to detect a dark matter particle in the laboratory; and (ii) the presence or absence of such a disc constrains the merger history of our Galaxy. In this work, we develop a chemo-dynamical template to hunt for the accreted disc. We apply our template to the high-resolution spectroscopic sample from Ruchti et al. (2011), finding at present no evidence for accreted stars. Our results are consistent with a quiescent Milky Way with no $\simgt 1:10$ mergers since the disc formed and a correspondingly light `dark disc'. However, we caution that while our method can robustly identify accreted stars, our incomplete stellar sample makes it more challenging to definitively rule them out. Larger unbiased stellar samples will be required for this.
\end{abstract}


\section{Introduction}\label{sec:introduction}

In our current cosmological model ($\Lambda$ Cold Dark Matter; $\Lambda$CDM), galaxies are expected to experience many mergers over their lifetimes \citep[e.g.][]{white1978}. The substructures and streams now observed in the faint stellar halos that surround galaxies is strong confirmation of this basic idea \citep[e.g.][]{2007ApJ...671.1591I,2010AJ....140..962M,2014A&A...562A..73G}. However, it remains to be seen whether the expected number and mass of mergers predicted in theory agrees quantitatively with the observations \citep[e.g.][]{2005ApJ...635..931B,bailin2014}. Our own Galaxy is a key data point in this endeavour since we can study it in greater detail than any other. The proximity of stars in the Solar neighbourhood allows us to measure their individual phase space positions and chemistry, opening up the possibility of empirically tracing their birthplaces -- so-called Galactic Archaeology \citep[e.g.][]{eggen1962,freeman2002}.

The Milky Way stellar halo is a natural first place to look for accretion debris since the orbit time of stars in the outer halo is long and substructure can remain for over a Hubble time without fully phase mixing (see \S\ref{sec:theory}). Observational evidence for such `detritus' continues to grow as Galactic surveys improve in sky-coverage and depth \citep[e.g.][]{1999Natur.402...53H,ibata2002,ibata2003,2006ApJ...642L.137B}. However, the phase mixing time is shortest for the most massive mergers, meaning that the stellar halo is most sensitive to low mass accretion events. Indeed, some of the most beautiful stellar streams come from globular clusters and must have formed over billions of years \citep[e.g.][]{2001ApJ...548L.165O,2006ApJ...639L..17G}. 

Like the stellar halo, the Galactic disc can also be expected to retain traces of past accretion. The key difference is that the disc is most sensitive to massive mergers. There are two key signatures that we can look for. Firstly, massive mergers perturb and heat the disc causing it to flare and warp \citep[e.g.][]{read2008,kazantzidis2008,villalobos2008}, driving the radial migration of stars \citep[e.g.][]{minchev2013,minchev2014}. For the most massive mergers, the disc may even be destroyed \citep[e.g.][]{toomre1977,barnes1992,steinmetz2002}. Secondly, massive prograde satellites are preferentially dragged towards the disc plane by dynamical friction, an effect that we call `dynamical friction plane dragging' \citep{1986ApJ...309..472Q,1989AJ.....98.1554L,1996ApJ...460..121W,2003ApJ...597...21A}. There, as they are torn apart by tidal forces, they deposit their stars and dark matter in an accreted stellar/dark disc \citep{1989AJ.....98.1554L,read2008,2009MNRAS.397...44R,2009ApJ...703.2275P,2010JCAP...02..012L,pillepich2014}. The accreted disc is not expected to contribute significantly to the Milky Way thin or thick discs \citep[e.g.][]{1996ApJ...460..121W,gilmore2002,wyse2006,read2008}. Indeed, from chemistry alone, it has long been known that the thick disc \citep{gilmore1983} cannot have formed from the detritus of accreted satellites alone \citep[e.g.][]{gilmore1989,ruchti2011mpd,minchev2013}. However, the accreted disc remains interesting for two reasons. Firstly, the accreted dark disc boosts and alters the signals expected in particle dark matter search experiments \citep{bruch2009,2009PhLB..674..250B,2010JCAP...02..012L}. Secondly, the stellar accreted disc is a direct probe of late massive mergers after the Milky Way disc formed. 

In this paper, we set out to find or constrain the Milky Way's accreted stellar disc and thereby probe its dark matter disc. The first work to look for such accreted disc structures was \citet{2006MNRAS.365.1309H}, using the Geneva-Copenhagen Survey (GCS) data \citep{2004A&A...418..989N}. They used simulations to motivate their search, looking for over-densities in the specific energy $E$ and the $z$-component of the specific angular momentum $J_z$ of stars (both conserved quantities for orbits in static axisymmetric potentials). They identified three coherent groups of stars that were additionally clumped in stellar age and metallicity, attributing each to a merger remnant, the most massive of which had a stellar mass of $\sim 4\times 10^8$\,M$_\odot$. All three satellites uncovered in their work were consistent with early accretion before $z\sim1$, with little merger activity thereafter.

In this paper, we also hunt for evidence for massive accretion events in the Milky Way disc, but taking a different approach to \citet{2006MNRAS.365.1309H}. Searching for over-densities in phase space is greatly facilitated by either a complete and unbiased sample of stars with 6D phase space data and chemistry, or a sample for which such biases are well understood. This is only currently available for data within $\sim 40$\,pc of the Sun \citep{2004A&A...418..989N}. To avoid this problem and thereby extend our search over a much larger volume, we introduce a novel chemo-dynamic template for accreted disc stars in the Milky Way. We show that our template can be applied to even incomplete stellar samples with an unknown selection function to robustly determine the presence of accreted stars. However, the converse is not true. If no accreted stars are found, an unknown stellar selection function makes it more challenging to conclusively rule out massive accretion events. We apply our template to a high-resolution spectroscopic sample of stars with full 6D phase space data \citep{ruchti2011mpd,ruchti2013} to hunt for evidence of massive accretion events in the Milky Way.

This paper is organised as follows. In \S\ref{sec:theory}, we briefly review the relevant background theory necessary for building our `accreted disc' chemo-dynamic template. In \S\ref{sec:thetemplate}, we describe our chemo-dynamic template. In \S\ref{sec:obsdata}, we describe the observational data. In \S\ref{sec:results}, we compare these data to our template to place constraints on massive accretion events in our Galaxy. Finally, in \S\ref{sec:conclusions}, we present our conclusions.

\section{Theory}\label{sec:theory} 

In this section, we briefly review the relevant theory necessary for building our `accreted disc' chemo-dynamic template. In \S\ref{sec:phasemix}, we show that the stellar halo is most sensitive to small substructures, while the disc is most sensitive to massive ones. We then define {\it accreted disc stars} as distinct from {\it accreted halo stars}. The former come from the very most massive mergers that are more metal rich and that suffer dynamical friction during interactions with the Galactic disc that alters the resultant kinematics of the accreted stars.

\subsection{What are accreted disc stars?}\label{sec:phasemix}

The stellar halo is a living fossil record of past accretion events. Consider a small substructure -- that can be a star cluster or small dwarf galaxy -- that has recently tidally disrupted and become unbound\footnote{Note that the timescale for unbinding is different from the phase mixing timescale we discuss here and could be significantly longer depending on the density of the satellite and its orbit \citep[e.g.][]{2006MNRAS.366..429R}.}. Let us assume for simplicity that it was, prior to disruption, moving on a circular orbit of radius $d$ about a spherical host galaxy with enclosed mass $M$. Its stars will then have a spread of orbital periods $\Delta T$ that is dictated by the velocity dispersion $\sigma$ (and therefore the mass) of the progenitor system: 

\begin{equation} 
\Delta T \sim \frac{6\sigma}{v} t_{\rm orb} 
\label{eqn:deltaT}
\end{equation} 
where $t_{\rm orb}$ is the orbital time for a circular orbit at $d$\footnote{Note that we obtain exactly the same scaling if we assume instead that stars escape at zero velocity from the tidal radius; see Appendix \ref{alternativephasecalc}.}.

We can crudely estimate the phase mixing time as $t_{\rm mix} \sim n t_{\rm orb}$, where $n$ is the number of orbits over which $n\Delta T \sim t_{\rm orb}$. Thus, we derive: 

\begin{equation}
t_{\rm mix} \sim \frac{v}{6\sigma}t_{\rm orb}
\end{equation}
We can now calculate what this timescale is for small star clusters or dwarf galaxies in the outer halo. Assuming the mass of the Milky Way inside a radius of 50~kpc is $\sim5\times10^{11}$\,M$_\odot$ \citep[e.g.][]{2002ApJ...573..597K}, we find that $t_{\rm orb} \sim 1.5$\,Gyr $\sim 0.1 t_{\rm Hubble}$, where $t_{\rm Hubble}$ is the age of the Universe.  Thus, globular cluster streams with $v/\sigma \sim 70$ will remain visible today, while massive mergers with $v/\sigma \sim 4$ (i.e. of mass ratio $\sim 1:16$) will mix away in less than an orbital time. If the Galactic potential is spherical or axisymmetric, we will still see these mergers as over-densities in specific energy $E$ and specific angular momentum $J_z$, since these are conserved quantities \citep[e.g.][]{2000MNRAS.319..657H}. However, large stellar samples with 6D data are required to detect such clumps over the background. Furthermore, any halo triaxiality will cause such structures to smear out also in $J_z$. 

While it is challenging to detect massive mergers via stellar kinematics alone, we might still hope to see evidence for them in the chemistry of accreted stars. Satellite galaxies form stars with a different efficiency than the Milky Way, resulting in distinct elemental abundance patterns \citep[e.g.][]{hendricks2014}. We discuss this in detail in \S\ref{sec:thetemplate}. Furthermore, these more metal rich accreted stars should also show distinct kinematics as compared to metal poor halo stars. This is because of dynamical friction plane dragging \citep{1986ApJ...309..472Q,1989AJ.....98.1554L,1996ApJ...460..121W,2003ApJ...597...21A}. All merging satellites experience dynamical friction that owes to momentum exchange between the in-falling satellite and some background stars and/or dark matter: 
\begin{equation} 
M_{\rm sat} \dot{\bf v} \simeq C \frac{\rho M_{\rm sat}^2}{v^3}{\bf v} 
\label{eqn:friction}
\end{equation} 
where $M_{\rm sat}$ is the mass of the satellite; $\dot{\bf v}$ is the deceleration due to dynamical friction; $\rho$ is the background density (i.e. stars, gas, dark matter etc.); $C$ is some constant of proportionality; and $v = |{\bf v}|$ is the velocity of the satellite relative to the background\footnote{Apart from especially resonant situations, the above formula that owes to \citet{1943ApJ....97..255C} works remarkably well \citep[e.g.][]{2006astro.ph..6636R}.}.

There are three key points to note from equation \ref{eqn:friction}. Firstly, the force is proportional to the mass of the satellite $M_{\rm sat} \sim \sigma^2 r_{\rm sat}/G$. For the Milky Way, it only becomes significant for $M_{\rm sat} \simgt 10^{10}$\,M$_\odot$ and therefore for mergers with a mass ratio greater than 1:100 -- exactly the sort of massive mergers we are interested in here \citep[e.g.][]{lux2010}. Secondly, the force is also proportional to the background density $\rho$. This rises significantly if the satellite passes through the disc, leading to enhanced friction there. Over several orbits, this causes the satellite to be dragged down towards the disc plane. If this happens on a time-scale faster than the tidal disruption time, then the satellite will deposit its remaining stars and dark matter in an accreted disc-like structure. This defines what we call from here on {\it accreted disc stars}. Thirdly, notice that the frictional force also scales as $1/v^2$ where $v$ is the velocity of the satellite {\it relative} to the background. Thus, satellites that move on prograde orbits with respect to the disc have $v \sim 0$ and experience significantly enhanced friction as compared to retrograde satellites that have $v \sim 400$\,km/s. For this reason, we expect disc plane dragging to act preferentially on prograde satellites, producing accreted discs that typically co-rotate with the host galaxy \citep[e.g.][]{read2008,2009MNRAS.397...44R}. Exceptions to this rule are possible, of course, if a retrograde satellite is particularly massive or if it is particularly aligned with the disc plane in the first place. 

\subsection{Statistics for accreted discs in $\Lambda$CDM}\label{sec:accstat}

Assuming isotropic accretion, the probability of a merger within an angle $0 <\theta < \pi/2$ to the disc plane is given by $P(<\theta) = \sin\theta$. However, this does not distinguish between prograde or retrograde mergers. If we separate out these two possibilities, then the probability becomes instead $P(<\theta) = \sin^2(\theta/2)$ with $0 < \theta < \pi$, where retrograde mergers correspond to $\pi/2 < \theta < \pi$. The likelihood of a prograde merger within $30^\circ$ of the disc is then rather small ($\sim 7\%)$. However, this does not account for the disc-plane dragging effect, nor cosmological accretion that often occurs along filaments and is not isotropic. Using three hydrodynamic cosmological simulations of Milky Way mass galaxies, \citet{2009MNRAS.397...44R} found that the fraction of accreted stars in the solar neighbourhood (defined as $7 < r < 9$\,kpc; $|z| < 2.1$\,kpc) ranged from 5\% to 40\% depending on the merger history of the galaxy, with the former being a particularly quiescent example and the latter having a near 1:1 mass ratio planar merger at redshift $z \sim 1$. The accreted stars were found to be moving on preferentially prograde orbits, particularly in the two simulations with significant mergers. The most quiescent case showed only a slight bias to prograde motion in the accreted star population, as expected given that the number of massive mergers was small, with correspondingly little disc-plane dragging. More recent higher resolution simulations are consistent with these numbers \citep[e.g.][]{pillepich2014}.

The above simulations demonstrate that once massive mergers occur, they will deposit prograde material into the disc plane. Conversely, if such accreted material is not present in the disc, this implies that the Milky Way must have been relatively quiescent. The statistics of minor mergers for Milky Way mass dark matter haloes were calculated in \citet{read2008} and \citet{2008ApJ...683..597S}. The former study used four high resolution simulations previously presented in \citet{DiemandMooreStadel2004a}; the latter used some 17,000 haloes simulated at lower resolution. Both studies were in excellent agreement, finding that 1:10 mass ratio mergers are common. Over the last $\sim 10$\,Gyrs, $\sim 95\%$ of Milky Way-sized haloes accreted at least one object more massive than the Milky Way disc. A `quiescent' Milky Way devoid of such massive mergers is then relatively rare, occurring 5-10\% of the time. 

\section{A template for accreted disc stars}\label{sec:thetemplate}

In this section, we describe our chemo-dynamic template that we will use to hunt for accreted disc stars. We build the chemical template by empirically comparing stars from nearby dwarf galaxies (that have a wide range of star formation histories) with Milky Way disc stars (both thin and thick) and we build our kinematic template by using previously published collisionless simulations of disc-satellite accretion events presented in \citet{read2008}.  We show that by themselves, neither the elemental abundances nor the kinematics of stars provide enough information to confidently identify accreted disc stars.  Thus, we require a combined chemo-dynamic template. 

\subsection{The chemical template}\label{sec:chemtemplate}

\begin{figure*}
\centering
\includegraphics[height=0.25\textwidth]{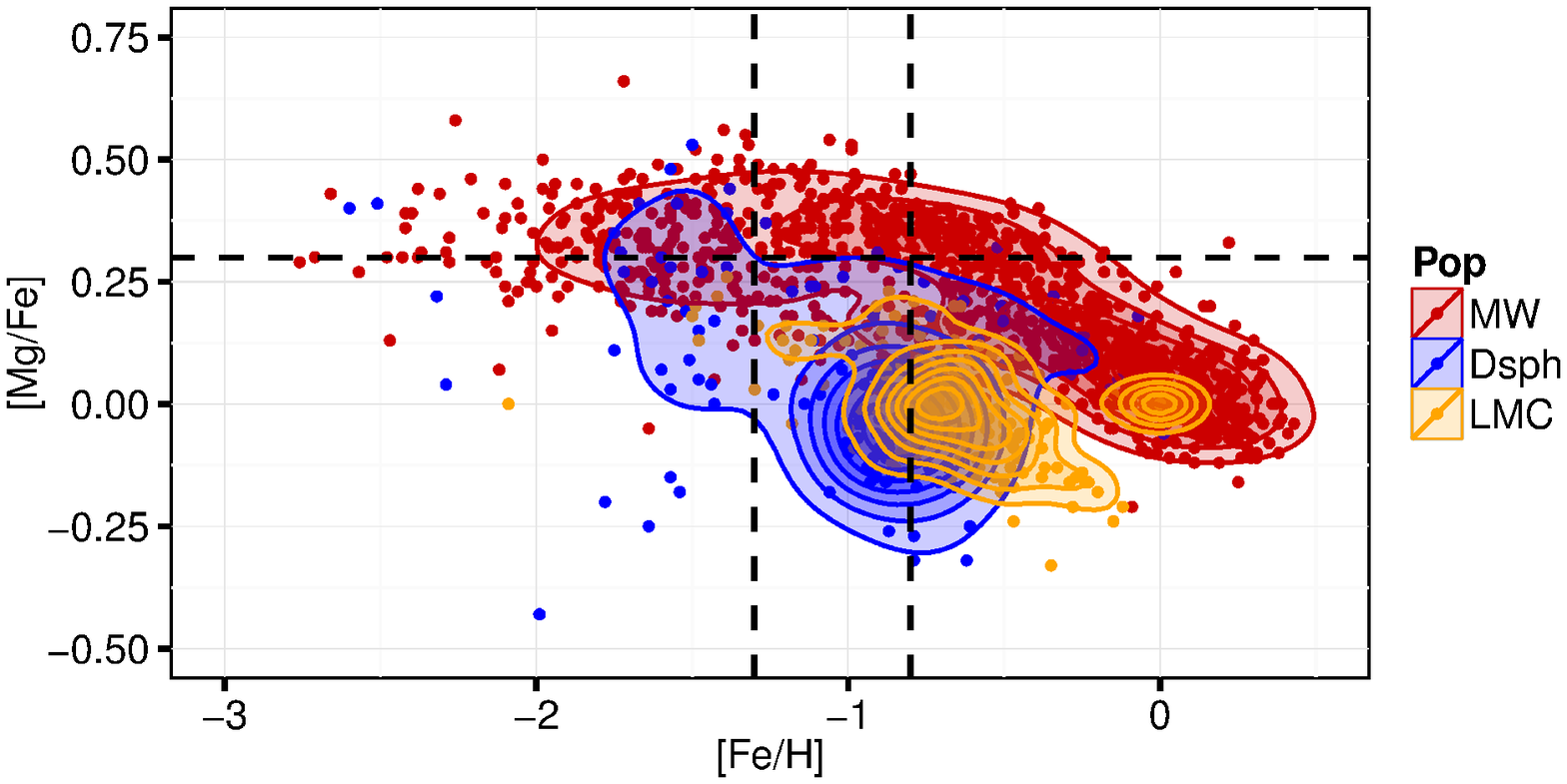}
\includegraphics[height=0.25\textwidth]{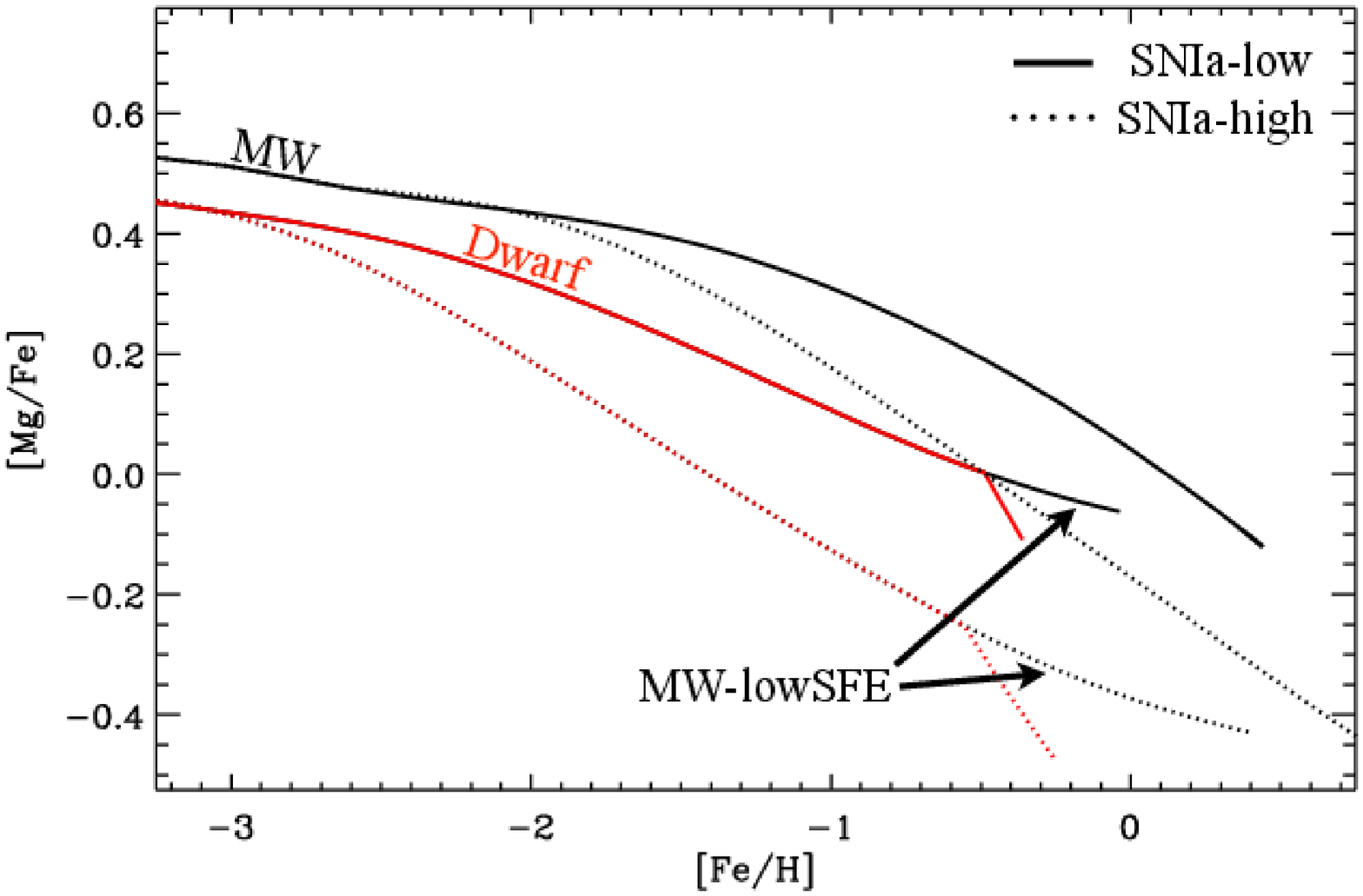}
\caption{{\bf Left:} Location of Milky Way stars (red); the LMC (orange); and giant stars in the Sagittarius, Carina, and Fornax dwarf galaxies (blue) in the \MA\ plane. Data compiled from \citet{bensby2014, ruchti2013, nissen2010,vanderswaelmen2013,carretta2010,lemasle2012,letarte2010}.  For $\feh>-1.3$ (vertical dashed lines) the satellite galaxies have [Mg/Fe] ratios typically less than $\sim0.3$~dex (horizontal dashed line), and are lower than the typical Milky Way star at similar metallicities.  {\bf Right:} A simple chemical evolution model to explain these data (see \S\ref{sec:chemmod} for details). We show results for stars in the \MA\ plane after 8\,Gyrs of star formation with an assumed exponentially declining gas inflow. We plot two models of different mass: $10^9M_{\odot}$ (Dwarf; red); and $10^{12}M_{\odot}$ (MW; black). The dwarf model assumes a low star formation efficiency $\nu = 0.1$/Gyr; we consider both $\nu=0.1$/Gyr and 1/Gyr for the MW model. In both cases, we plot results for two different SNIa rates (solid and dotted lines, as marked).}
\label{fig-dga} 
\end{figure*}

It is well known that dwarf galaxies present in the Local Group have $\afe$ ratios that are typically lower than those found in Milky Way stars at metallicities, $\feh>-1.5$ \citep{gilmore1991,tolstoy2009,hendricks2014}. Furthermore, the Large and Small Magellanic Coulds (LMC and SMC) have also been shown to have lower $\alpha$-enhancement than the Milky Way \citep[e.g.][]{dufour1984,russell1988,vanderswaelmen2013}. We show this for a compilation of data from the Milky Way disc and dwarf satellites in Figure \ref{fig-dga}a. In this figure, it is clear that for metallicities above $\feh=-1.3$, the majority of the stars in satellite galaxies have [Mg/Fe] ratios below $+0.3$~dex.

\subsubsection{The chemical evolution model}\label{sec:chemmod}

To understand {\it why} satellite galaxies typically have lower $\alpha$-abundances, we use a simple model to investigate the chemical evolution of galaxies as a function of their mass and star formation efficiency. The chemical evolution calculation is performed by means of the
multi-zone model of \citet[][PM04 hereafter]{pipino2004} in which we adopt a power law IMF with $x=1.35$
\citep{salpeter1955} with a star formation rate proportional
to the gas mass via a constant star
formation efficiency ($\nu$). We adopt an exponentially declining  
infall law for the rate at which primordial gas is accreted by the proto-galaxy.

For single low and intermediate mass stars ($0.8 \le M/M_{\odot} \le
8$), we use the yields from \citet{vandenhoek1997} as a function of metallicity. For massive stars ($M >8 M_{\odot}$), we adopt the yields of \citet{francois2004}. For SNIa, we use the yields of \citet[][model W7]{nomoto1997}. 
These are assumed to originate from C-O white dwarfs in binary systems
that have accreted material from a companion (the secondary), have reached
the Chandrasekhar mass, and have exploded via C-deflagration.

The model was originally developed to interpret the abundance pattern
of both present-day elliptical galaxies (PM04) and their high-redshift progenitors \citep[e.g.][]{pipino2011,pipino2013}.
In this paper, we use it as a tool to show the dependence of the $\mgfe$ ratio
on the depth of the potential well and the star formation efficiency $\nu$ in the Milky Way and its satellites.
We consider $\nu$ in the range $0.1-1$/Gyr with the infall time-scale fixed to 8\,Gyr. The model has not been tuned to reproduce the elemental abundance patterns found in the Milky Way or its neighbours. Rather, the model predictions are meant to offer a qualitative description of the abundance ratio trends with $\nu$ for galaxies like these.
We refer the reader to other works in the literature which adopt very similar numerical codes,
for a thorough and self-consistent interpretation of the chemical evolution of the Milky Way
\citep[e.g.][to mention some of the most recent versions]{chiappini1997,grieco2012,micali2013,brusadin2013} and of local dwarfs \citep[e.g.][]{lanfranchi2006,lanfranchi2010}.

In Figure \ref{fig-dga}b, we show results for two models with a fixed dark matter potential of $10^9M_{\odot}$ (Dwarf) and $10^{12}M_{\odot}$ (MW). We model the low mass case with a star formation efficiency of $\nu = 0.1$/Gyr; we adopt both $\nu=0.1$/Gyr and 1/Gyr for the higher mass case. For both models, we plot two curves showing results for low and high SNIa rates, varying the parameter $A$ (see equation 2 in PM04) from 0.05 to 0.18, respectively.

The figure clearly shows that the $\mgfe$ ratio slowly decreases with $\feh$, remaining almost flat until at some point (e.g. $\feh\sim-1$ for the MW case) the decrease becomes much more rapid.  This ``knee" in the curve is due to the onset of SNIa, which typically explode on longer time-scales.  This is known as the ``time-delay model" \citep{matteucci1986}. The position of the knee along $\feh$ thus depends on the SNIa rate (solid vs. dotted curves).

As indicated in Fig.~\ref{fig-dga}b, the star formation efficiency plays an important role in the resultant abundance patterns.  The low-mass model shows consistently lower $\mgfe$ ratios than the high-mass, high-$\nu$ model.  This is consistent with observations for $\feh>-2$ (see Fig.~\ref{fig-dga}a).  However, the high-mass model with low $\nu$ also shows lower $\mgfe$ ratios than that with high $\nu$. The fact that the \MA\ curves vary as a function of $\nu$ is not new \citep[][PM04]{matteucci1990,matteucci1994,lanfranchi2004}, what is important here is the illustration of how this links with the potential.  The star formation efficiency correlates both with the depth of the potential well, which affects the gas surface density, and the radius in the disc of the Milky Way at which stars form. 

A striking feature of Figure \ref{fig-dga}b is that the MW model with low $\nu$ sits on the exact same \MA\ curve as the dwarf galaxy model with identically low $\nu$.  The likeness between the two curves is, in part, due to the simplicity of our model.  However, the fact remains that stars formed in regions of low $\nu$ in the Milky Way could have similarly low $\alpha$-enhancements to stars formed in satellite galaxies. This poses a problem: stars with low $\mgfe$ in the solar vicinity could be accreted stars born in dwarf galaxies, or they could be `in-situ' stars born in the Milky Way disc. These latter `in-situ' stars could have formed in a variety of ways. They could simply be the low $\mgfe$ tail of a distribution of stars formed at a single epoch; they could have formed at the same location as the majority of thin disc stars but at a time when the surface density was lower; or they could have formed at radii $R>R_{\odot}$ (where the surface density is also lower) and later migrated inwards \citep[e.g.][]{sellwood2002,roskar2008a,roskar2008b,minchev2012}. 

Recent observational evidence suggests that stars in the outer disc of the Milky Way do indeed have lower $\mgfe$ ratios than those in the inner disc \citep[e.g.][]{bensby2011letter,anders2014,bergemann2014}. This could imply radial migration. Alternatively, it could be explained as a difference in the radial scale-length between distinct `thin' and `thick' stellar discs \citep[e.g.][]{bensby2011letter,cheng2012_2}. Either way, it is important to realise that radial migration complicates the interpretation of any such metallicity gradient observed today. Owing to radial migration, the abundance distributions of the disc, as observed today, may not reflect the distributions when the disc formed.

It is clear that to distinguish among the above possibilities, we require further information in the form of stellar kinematics and ages. We discuss the kinematic template for distinguishing accreted from `in-situ' stars, next.

\subsection{The kinematic template}\label{sec:kintemplate}

We use a subset of the simulations already published in \citet{read2008} for our kinematic template. These model collisionless minor mergers of satellites on cosmologically motivated orbits. We consider two satellite masses: an LMC-like galaxy with mass $2.4 \times 10^{10}$\,M$_\odot$ (stellar mass $7\times 10^8$\,M$_\odot$); and a `Large' LMC (LLMC) with mass $10^{11}$\,M$_\odot$ (stellar mass $3\times 10^9$\,M$_\odot$). These are placed on eccentric orbits at 10$^\circ$ and 20$^\circ$ to the disc plane of a Milky Way-mass host galaxy. We additionally consider one low inclination merger (10$^\circ$) at low eccentricity (LMC-eless). The simulations are evolved for several Gyr until the disc has settled into a steady state. In Figure \ref{fig-accretedsims}, we plot the distribution of `in-situ' disc stars (black contours) and accreted stars (blue contours); the red contours show the initial conditions for the in-situ disc stars, set up to mimic the Milky Way thin disc. We select only those stars that lie within a `Solar neighbourhood' patch ($8 < R < 9$\,kpc). We work in specific vertical energy-angular momentum space \EzJz: 

\begin{equation}
J_{\rm z} = R \cdot V_{\theta} \,\,\,\, ; \,\,\,\, J_{\rm c} = R_{\rm c} \cdot V_{\rm c}
\end{equation}
\begin{equation}
E_{\rm z} = \frac{V_{\rm z}^2}{2} + \Phi(R,Z) \,\,\,\, ; \,\,\,\, E_{\rm c} = \Phi(R,0)
\end{equation}
where $V_{\theta}$ and $V_z$ are the rotational and vertical velocity components with respect to the Galactic disc, respectively; $V_{\rm c}$ is the circular velocity:
\begin{equation}
V_{\rm c} = \sqrt{R_{\rm c} \cdot \left. \frac{\partial\Phi}{\partial R}\right|_{R=R_{\rm c}}}.
\end{equation}
and $R_{\rm c}$ is the radius of a planar circular orbit with the same specific energy $E$ as the star. This is found by minimising: 
\begin{equation}
\left| \frac{V_{\rm c}^2}{2} + \Phi(R_{\rm c},0) - E \right|,
\end{equation}
The advantage of using \EzJz\ space is: (i) these are integrals of motion for a static axisymmetric potential; and (ii) they are less sensitive to the choice of potential $\Phi$ than the orbital eccentricity $e$ and maximum vertical excursion $z_{\rm max}$ that are sometimes used instead. Circular orbits have $J_z/J_c = E_z/E_c = 1$. 

Firstly, notice in Figure~\ref{fig-accretedsims}a that the initial `thin disc' of stars occupies the region to the very top right of \EzJz\ space (red contours), corresponding to stars on nearly circular orbits in the disc plane. By contrast, after the merger, the heated stars (black contours) extend to lower vertical energy and lower angular momentum. The accreted stars occupy a space of even lower $E_z$ and $J_z$ (blue contours), clearly distinct from the heated stars. However, this depends on the orbit of the progenitor. In Figure~\ref{fig-accretedsims}c, we plot a lower eccentricity merger that produces accreted material that nearly perfectly overlaps with the heated stars. Such an orbit is cosmologically unlikely, when considering the mean of orbits extracted from ``dark matter only" cosmological simulations \citep[e.g.][]{read2008}. However, including stars and gas in such simulations could circularise satellite orbits leading to exactly these type of lower eccentricity mergers \citep[e.g.][]{1986ApJ...309..472Q}. Thus, we cannot conclude from the stellar phase space information alone that stars were accreted onto the disc.

\subsection{The chemo-dynamic template}\label{sec:chem-dyn-template}

\begin{figure*}
\begin{minipage}[b]{.5\linewidth}
\centering\begin{overpic}[scale=0.5]{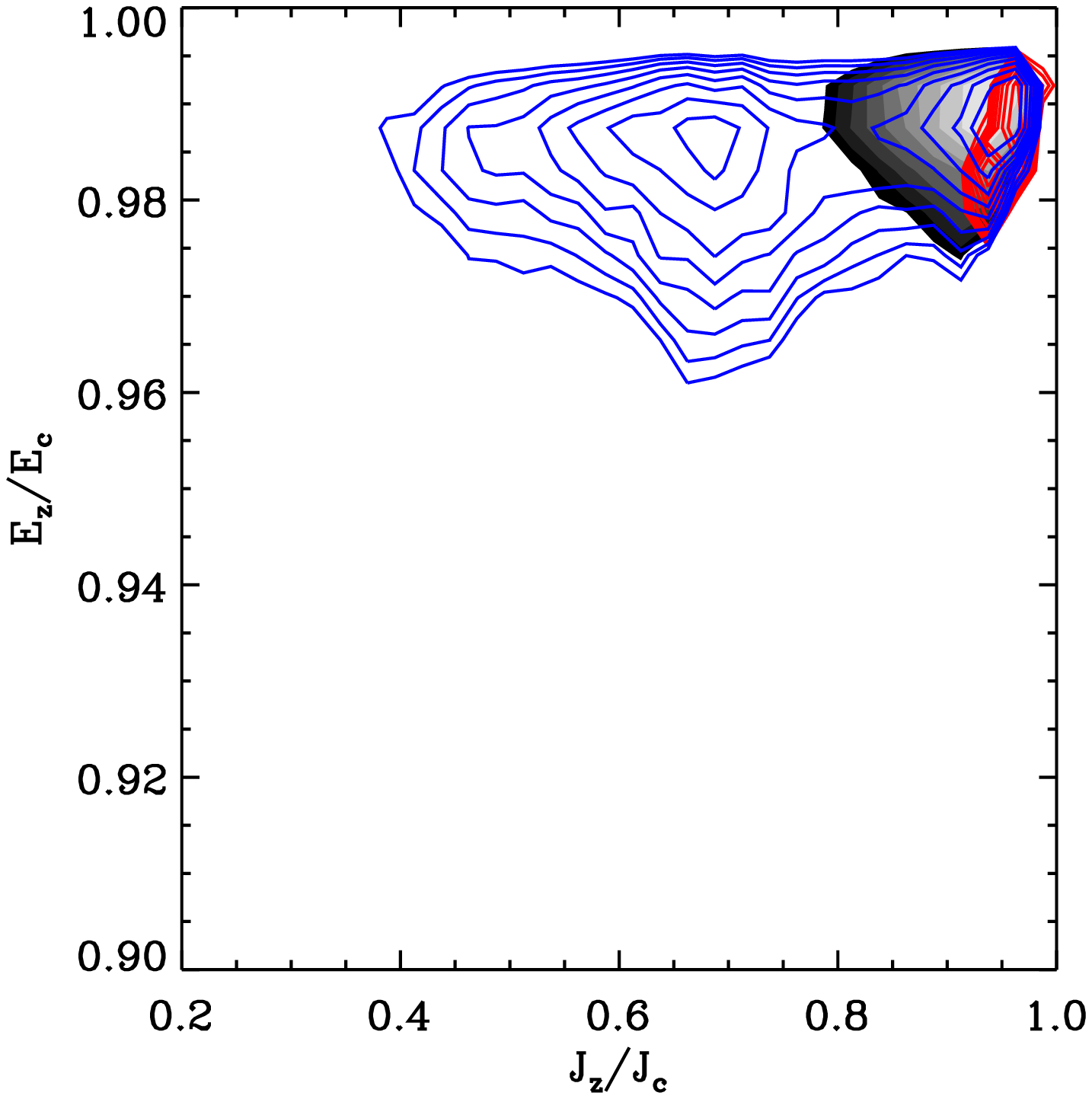}\put(20.5,95){a) LMC-10$^\circ$}\end{overpic}
\end{minipage}%
\begin{minipage}[b]{.5\linewidth}
\centering\begin{overpic}[scale=0.5]{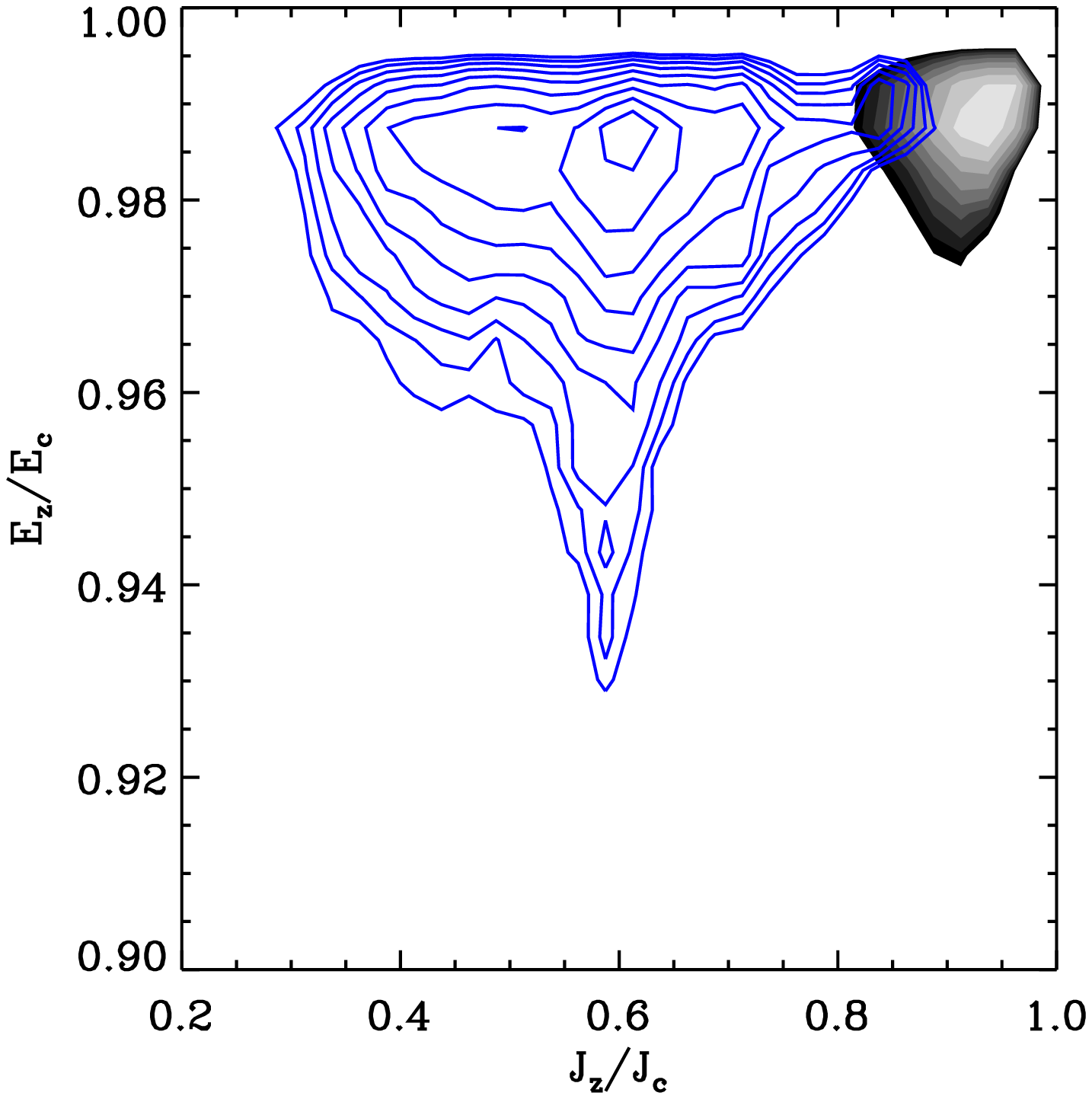}\put(20.5,95){b) LMC-20$^\circ$}\end{overpic}
\end{minipage}\\
\begin{minipage}[b]{.5\linewidth}
\centering\begin{overpic}[scale=0.5]{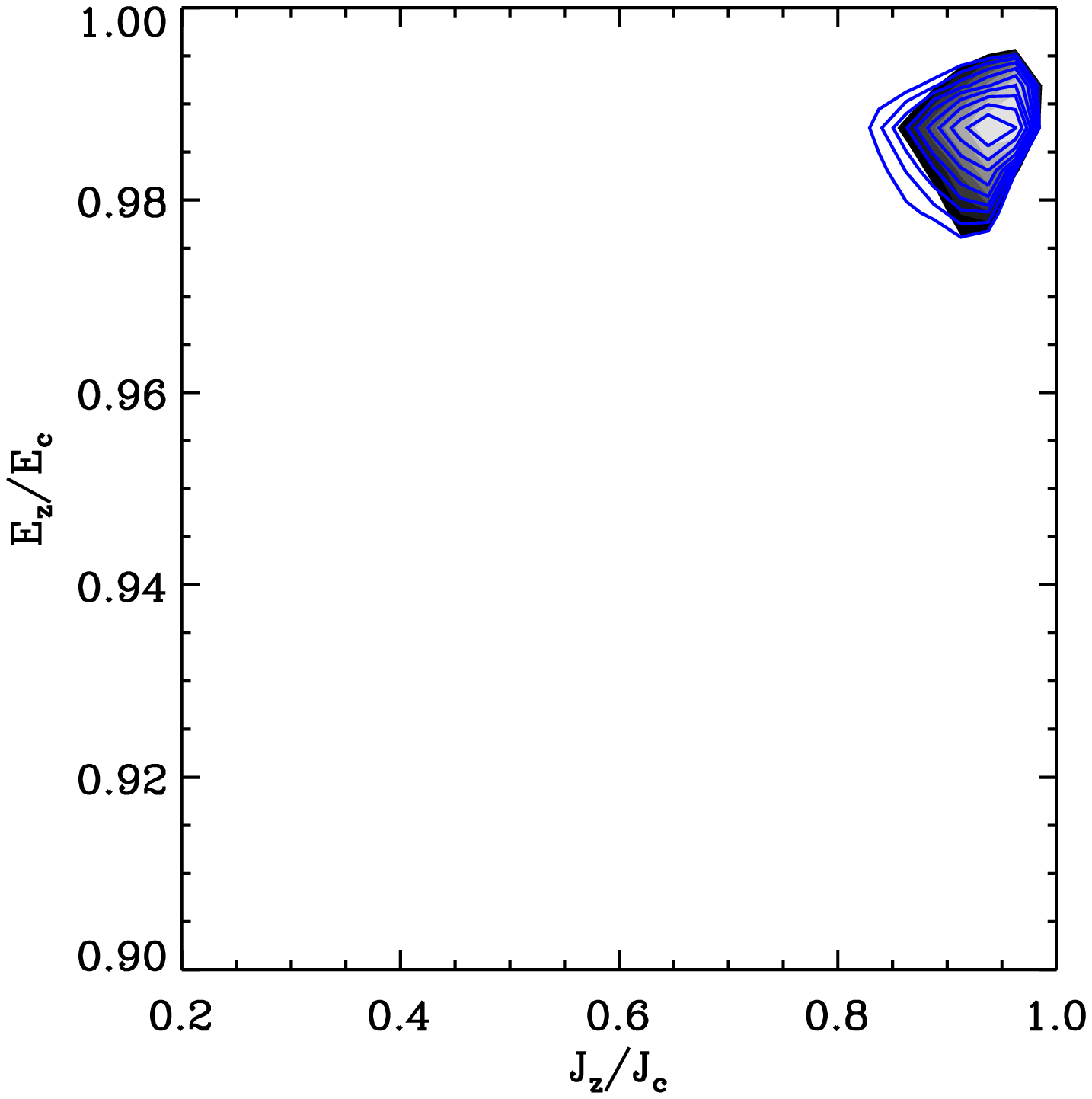}\put(20.5,95){c) LMC-eless-10$^\circ$}\end{overpic}
\end{minipage}%
\begin{minipage}[b]{.5\linewidth}
\centering\begin{overpic}[scale=0.5]{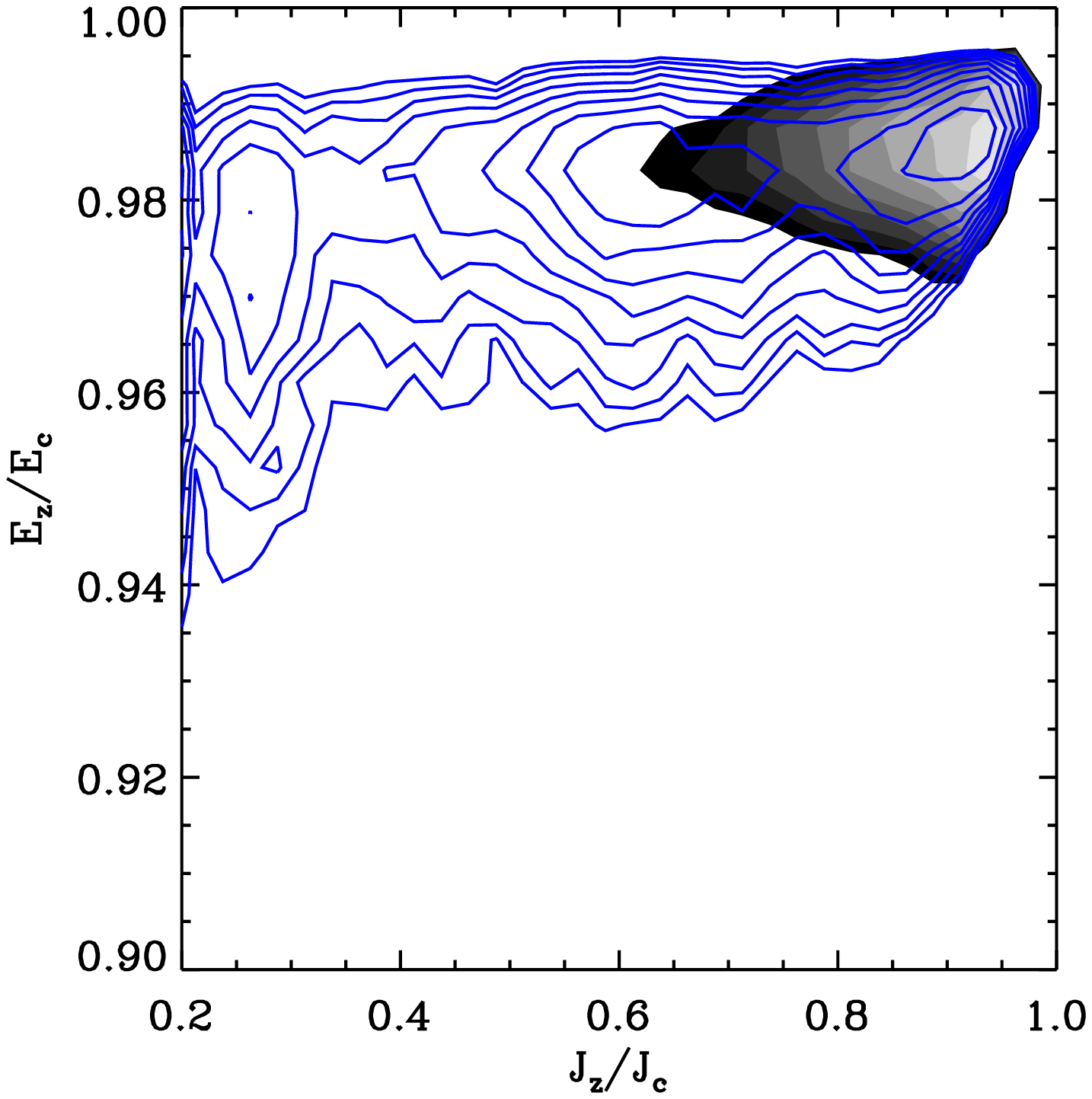}\put(20.5,95){d) LLMC-10$^\circ$}\end{overpic}
\end{minipage}
\caption{Solar neighbourhood ($8 < R < 9$\,kpc) heated versus accreted stars in a range of Milky Way minor merger simulations, projected into the vertical energy-vertical angular momentum plane: \EzJz.  Note that stars that travel high above the disc corresponds to low $E_z/E_c$; stars with high eccentricity correspond to low $J_z/J_c$. The contours show: initial condition in-situ stars in the disc (red; panel a only); heated in-situ stars (grey/black); and accreted stars (blue). The top two panels explore the effect of orbital inclination, showing a 10$^\circ$ (a) and 20$^\circ$ (b) `LMC'-like merger with the Milky Way.  The bottom two panels show results for a low eccentricity merger (c), and for a very massive ($\sim 4 \times$\,LMC) 10$^\circ$ merger (d).}
\label{fig-accretedsims}
\end{figure*}

Both the chemical and kinematic templates show degeneracies between accreted and in-situ stars in the disc. However, combining the two allows us to simultaneously break the degeneracies in both. Stars that migrate from the outer disc can have identical chemistry to stars born in satellite galaxies (see Figure \ref{fig-dga}). However, they will have thin disc-like kinematics, moving on orbits too circular to be consistent with even a low-eccentricity accretion event \citep[e.g.][and see Figure \ref{fig-accretedsims}c]{sellwood2002,binney2013,sellwood2014}. In this way, our {\it chemo-dynamic} template allows us to distinguish accreted stars from `in-situ' stars in the disc. Based on our results in Figures \ref{fig-dga} and \ref{fig-accretedsims}, we define `accreted stars' as having: $\mgfe < 0.3, \jzjc < 0.8$ and $\ezec < 0.97$. 

\section{Observational Data}\label{sec:obsdata}

To hunt for massive accretion events, we require a sample of stars with precisions in elemental abundances $\simlt0.1$~dex in order to distinguish populations according to our chemical template \citep[see also][]{lindegren2013}. We also require stars with a broad spread of kinematic properties (stars selected to lie only in the disc plane on circular orbits will necessarily select against accreted disc stars). We have thus chosen the high-resolution spectroscopic sample of \citet{ruchti2011mpd,ruchti2013} for our study.    

The \citet{ruchti2011mpd} data set consists of 319 sub-giant, red giant branch (RGB), and horizontal branch stars originally selected for high-resolution spectroscopic observations from the Radial Velocity Experiment Survey \citep[RAVE,][]{steinmetz2006} to study a large sample of metal-poor stars with thick disc-like kinematics.  Note, however, that this selection was based on distances and 3D space motions derived using the stellar parameters in the first and second data releases from RAVE.  Thus, systematics in the RAVE stellar parameters \citep[see][]{ruchti2011mpd,serenelli2013} resulted in the final sample containing a mixture of both disc and halo stars.  In \citet{ruchti2013}, more accurate stellar parameters were derived using a novel methodology, which included on-the-fly non-local thermodynamic corrections. Further, distances to the 286 sub-giant and RGB stars were computed from fits to evolutionary tracks using Bayesian techniques \citep[see][for more details]{serenelli2013}.  We apply our analysis to these 286 stars, hereafter referred to as R13. 

\subsection{Coping with selection bias}\label{sec-bias}

\begin{figure}
\centering
\includegraphics[height=0.45\textwidth]{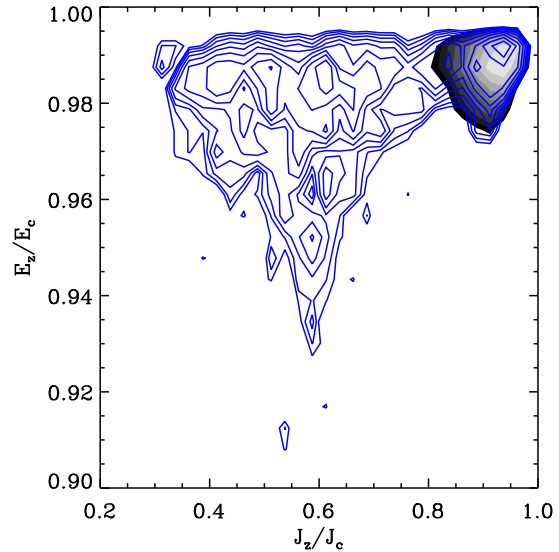}
\caption{The effect of the \citet{ruchti2011mpd} selection function on the distribution of heated `in situ' stars (black/grey) and `accreted' stars (blue). We show results in the \EzJz\ plane for the LMC-$20^{\circ}$ merger simulation, assuming a Solar neighbourhood mass fraction in accreted stars of $\eta = 0.4$. Although the relative number density of the two populations is clearly different from that in Figure~\ref{fig-accretedsims}b, the separation between the two populations is still clear. The vast majority of stars identified as `accreted stars' by our chemo-dynamic template are indeed accreted stars, rising to 100\% for our template cuts: $\jzjc < 0.8$ and $\ezec < 0.97$.}  
\label{fig-bias}
\end{figure}

Given the expense of taking high resolution spectra for stars, stars with an excellent measure of abundance and metallicity are not numerous. Furthermore, their observational selection function is typically complex, depending on lower resolution (and typically systematically biased) surveys for pre-selection, and other confounding factors like the weather on a given observing night \citep[e.g.][]{edvardsson1993,bensby2003,ruchti2011mpd,bergemann2014}. To estimate what effect the selection function for the R13 data sample has on our chemo-dynamic template, we use the simulations described in \S\ref{sec:kintemplate}. We proceed as follows:

\begin{enumerate} 

\item We paint metallicity and abundance on the simulation star particles, assuming Gaussian distributions. For the in-situ stars, we assume a mean of $\feh=-0.6$ with dispersion of $0.25$~dex and a mean [Mg/Fe] of $0.35$~dex with a dispersion of $0.1$~dex; for the accreted stars, we assume the same metallicity distribution, but draw the abundances from a Gaussian with mean ${\rm [Mg/Fe]}=0$ and dispersion $0.1$~dex. 

\item We combine the accreted and in-situ stars assuming a normalisation of $\eta \equiv M_a /  M_i = 0.02, 0.1$ and $0.4$ (where $M_a$ and $M_i$ are the total mass of accreted and in-situ stars in the Solar neighbourhood, respectively).

\item We mimic the \citet{ruchti2011mpd} selection function by: (i) cutting away all stars with $[{\rm Fe/H}] > -0.5$; and (ii) accept/rejecting stars by drawing from a `thick disc' velocity distribution function: 

\begin{equation} 
f = \exp\left(-\frac{v_R^2}{\sigma_R^2}\right)\exp\left(-\frac{(v_\phi-\overline{v}_\phi)^2}{\sigma_\phi^2}\right)\exp\left(-\frac{v_z^2}{\sigma_z^2}\right)
\end{equation}
with $\sigma_R = 63$\,km/s; $\sigma_\phi = 39$\,km/s; $\overline{v}_\phi = 159$\,km/s; and $\sigma_z = 39$\,km/s.

\item Finally, we assign each star to an `in-situ' and `accreted' population using the definition in \S\ref{sec:chem-dyn-template}: $\mgfe < 0.3$ (`accreted'); $\mgfe > 0.3$ (`in-situ'). This hard cut necessarily means that some accreted stars are misclassified as in-situ and visa-versa. 

\end{enumerate} 
We plot the resulting distribution of stars for the LMC-20$^\circ$ simulation in \EzJz\ space for $\eta = 0.4$ in Figure \ref{fig-bias}. Notice that the density contours of `accreted' versus `in-situ' stars is now very different from that plotted in Figure \ref{fig-accretedsims}. Since we do not accurately know the \citet{ruchti2011mpd} selection function, this means that we cannot hope to attribute quantitative meaning to the relative numbers of `accreted' versus `in-situ' stars, nor their density distributions in \EzJz\ space. However, stars that inhabit the region defined by our chemo-dynamic template ($\mgfe < 0.3, \jzjc < 0.8$ and $\ezec < 0.97$) are all genuine accreted stars. Thus, we can be confident that our template will correctly identify accreted disc stars, even if it is more challenging to determine the number density of such stars.  

\subsection{The effect of sampling}\label{sec:sampling} 

\begin{figure*}
\centering
\includegraphics[width=\textwidth]{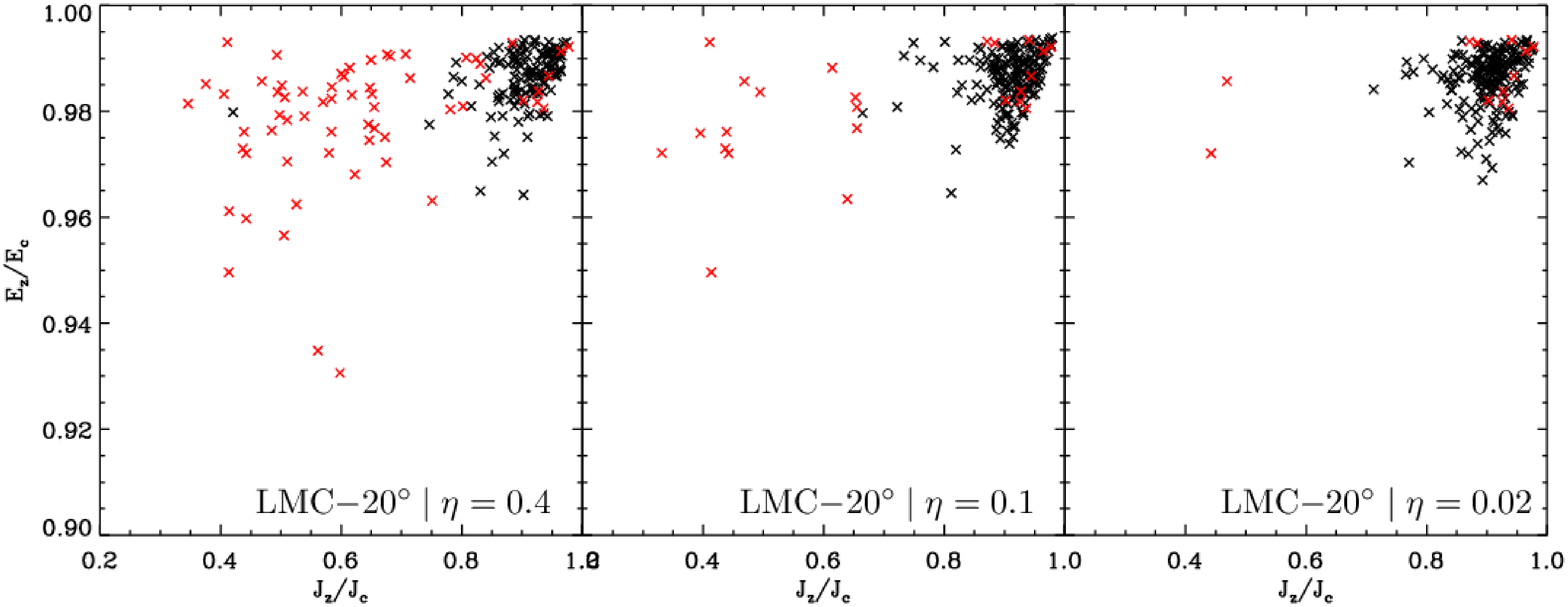}
\caption{The effect of the \citet{ruchti2011mpd} selection function combined with poor sampling. We draw 286 stars from the full distribution and then cut on [Fe/H] > $-0.8$ consistent with that expected for massive mergers. In this plot, the `in-situ' stars (black data points) and `accreted stars' (red data points) are both defined using our chemo-dynamic \mgfe\ cut. We show results in the \EzJz\ plane for the LMC-$20^{\circ}$ merger simulation. From left to right, we assume that the accreted stars comprise 40\%, 10\% and 2\% by mass of the in-situ population. Even with a normalisation of just 2\%, we should detect of order 2 accreted stars with our chemo-dynamic template.}
\label{fig-sampling}
\end{figure*}

Finally, we may worry that with just 286 stars (after all the selection cuts) as in the R13 sample we may not see the accreted disc simply because we would have to be very lucky to draw an accreted disc star from the full distribution. It is difficult to be quantitative about this because of the uncertainties in the \citet{ruchti2011mpd} selection function. However, we can use our simplified `selection effect' model from \S\ref{sec-bias} to explore the order-of-magnitude effect of poor sampling. In Figure \ref{fig-sampling}, we draw 286 stars from the biased distribution presented in Figure \ref{fig-bias}. We assume three different accreted to in-situ star fractions in the Solar neighbourhood: $\eta = 0.02, 0.1$ and $0.4$, and we cut stars on metallicity [Fe/H] > $-0.8$ consistent with what we expect for massive accretion events (see \S\ref{sec:thetemplate}). As can be seen, even if accreted stars make up just 2\% by mass of the in-situ population, and even after applying selection effects similar to those in \citet{ruchti2011mpd}, we can hope to see of order 2 accreted stars in our most metal rich bin using our chemo-dynamic template. This suggests that an absence of such stars really would imply a rather quiescent merger history for our Galaxy. It is, however, difficult to be much more quantitative than this; that requires a survey that is either complete or has a well-defined stellar selection function that can be modelled.

\section{Data Analysis}

\subsection{Mg abundances}

The abundance of Mg for each star was derived from the equivalent widths of \nion{Mg}{I} lines with the MOOG abundance analysis program \citep{sneden1973}, using one-dimensional, plane-parallel Kurucz model atmospheres \citep{castelli2004} and the stellar parameters derived in \citet{ruchti2013}.  The abundance analysis yielded an average precision in [Mg/Fe] of $0.04\pm0.02$~dex.   

\subsection{Space velocities}

We derived new space velocities for the R13 sample using the distances derived in \citet{serenelli2013} in combination with the proper motions given in the RAVE database and the radial velocities derived in \citet{ruchti2011mpd}.  The velocities were normalised to a Solar velocity with respect to the Local Standard of Rest (LSR), given by $(U_{\odot}, V_{\odot}, W_{\odot})=(14.00, 12.24, 7.25)$~\kmsec{} from \citet{schonrich2010} and \citet{schonrich2012}.  An additional $V_{\rm LSR}=220$~\kmsec{} was added to put the space motions in the Galactic rest frame.  The mean uncertainty on the space motions was less than 15\%.  

\subsection{Orbits}

The orbital information for each star was computed based on a three-component Galactic potential.  We assumed a \citet{hernquist1990} potential for the bulge,
\begin{equation}
\Phi_{\rm bulge} = -\frac{GM_{\rm bulge}}{r+c},
\end{equation}
and the \citet{miyamoto1975} potential for the disc:
\begin{equation}
\Phi_{\rm disc} = -\frac{GM_{\rm disc}}{\sqrt{R^2 + (a + \sqrt{z^2 + b^2})^2}}.
\end{equation}
Finally, we assumed a NFW dark matter profile \citep{navarro1996} for the halo:
\begin{equation}
\Phi_{\rm halo} = -\frac{GM_{\rm vir}}{\sqrt{R^2+z^2} \cdot d} \cdot \log\left(1 + \frac{\sqrt{R^2 + z^2}}{r_{\rm s}}\right)
\end{equation}
Masses and constants for each potential were adopted from \citet{gomez2010}, such that $V_{\rm c} = 220$~\kmsec{} at $8$~kpc from the Galactic Centre.

As described in \S\ref{sec:kintemplate}, we computed the specific vertical energy ($\ezec$) and vertical angular momentum ($\jzjc$) of each orbit.  In this way, we can directly compare the kinematics of the sample stars to the models.  Uncertainties in $\jzjc$ and $\ezec$ were propagated from the uncertainties in the space motions and positions.  This resulted in uncertainties of  $0.1\pm0.1$ in $\jzjc$ and $0.01\pm0.03$ in $\ezec$.  These are small enough to be able to distinguish between accreted stars and ``heated" disc stars according to our chemo-dynamical template (\S\ref{sec:chem-dyn-template}).

\section{Results}\label{sec:results}

\begin{figure}
\centering
\includegraphics[height=0.23\textwidth]{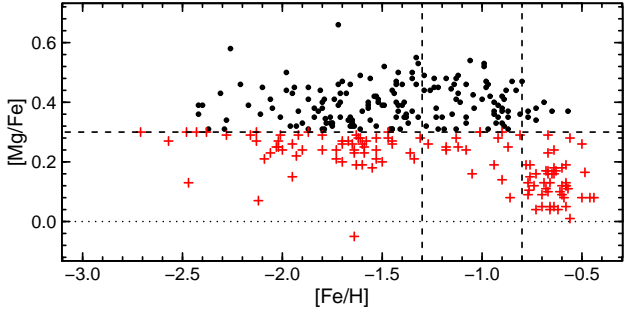}
\caption{The high-resolution spectroscopic sample, R13, plotted in the [[Mg/Fe],$\feh$] plane.  Stars are separated into {\it high-} and {\it low-} alpha regimes, with a cut at 0.3~dex in [Mg/Fe] (as defined in \S\ref{sec:chemtemplate}).  The data were divided into different metallicity bins shown by the vertical dashed lines.  The uncertainty in $\feh$ and $\mgfe$ are less than $0.1$~dex.  Subsequent plots are colour-coded in the same manner.}
\label{fig-afe}
\end{figure}

Based on the [Mg/Fe] ratios of satellite galaxies as compared to the Milky Way, shown in Figure~\ref{fig-dga}a, we separated the data into two regimes, defined by a cut at ${\rm [Mg/Fe]}=0.3$ (illustrated by the horizontal, dashed line in Figure~\ref{fig-dga}a).\footnote{Note that cutting at a lower ratio, e.g.  ${\rm [Mg/Fe]}=0.2$, does not affect our overall results.  See Appendix~\ref{alternativemgfe}.}  We consider stars that lie above this cut as {\it high-$\alpha$}, while those that lay below are {\it low-$\alpha$} and have elemental abundances similar to present-day dwarf satellite galaxies and the LMC.  We further divided the data into three metallicity bins, separated at $\feh=-1.3$ and $-0.8$~dex.  The low-metallicity cut was defined as the metallicity at which the satellite galaxies show [Mg/Fe] ratios distinct from the Milky Way (see Figure~\ref{fig-dga}a), while we added an additional cut at $\feh=-0.8$, since this has been shown to be the limiting metallicity of the thin disc of the Milky Way \citep[e.g.,][]{bensby2014}.  The R13 data, cut into the respective bins, are shown in Figure~\ref{fig-afe}.    

\begin{figure}
\centering
\includegraphics[height=0.6\textwidth]{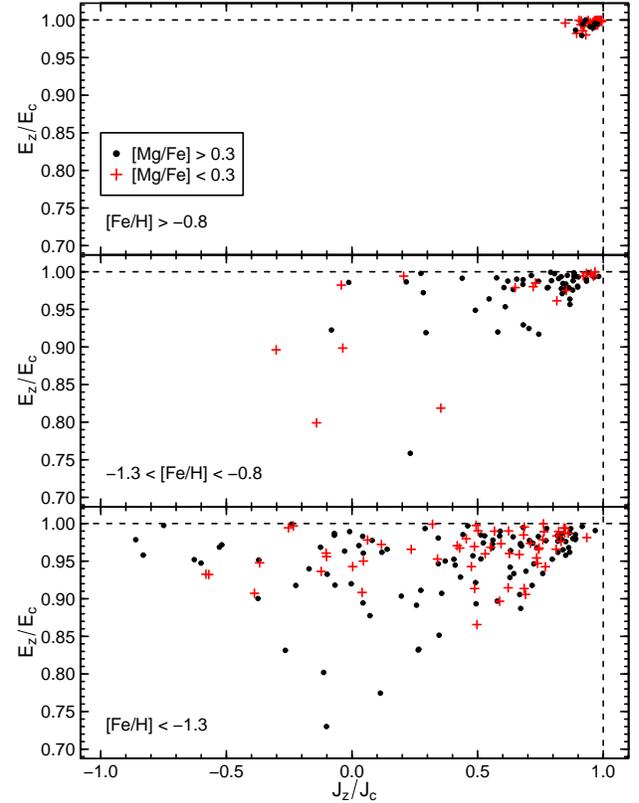}
\caption{The vertical component of the specific orbital angular momentum vs. the vertical component of the orbital energy for the R13 data set.  These can be directly compared to the LMC-merger simulations in Figure~\ref{fig-accretedsims}.  The symbols and colours are the same as those in Figure~\ref{fig-afe}.  The uncertainties are $0.1\pm0.1$ in $\jzjc$ and $0.01\pm0.03$ in $\ezec$.  {\it Top}: All stars with $\feh\geq-0.8$.  {\it Middle}: Stars with $-1.3\leq\feh<-0.8$.  {\it Bottom}: Stars with $\feh<-1.3$.}
\label{fig-ang}
\end{figure}

\begin{figure*}
\centering
\includegraphics[height=0.7\textwidth]{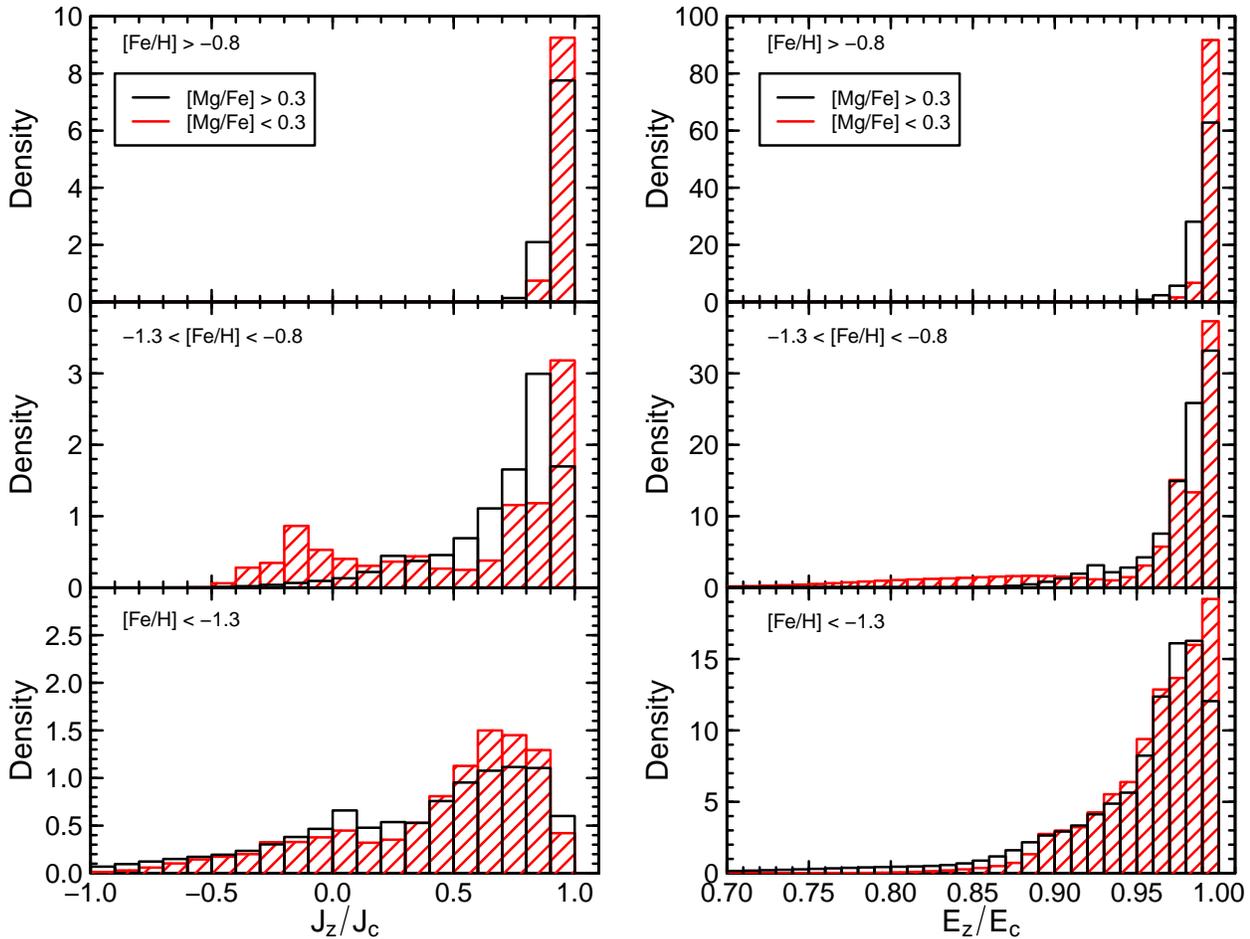}
\caption{Distributions of the vertical component of the specific orbital angular momentum and the vertical component of the orbital energy for the R13 data set.  Each distribution was constructed by summing the probability density function values within a given $\jzjc$ or $\ezec$ bin, and normalizing such that the total area equals unity.  The colours are the same as those in Figure~\ref{fig-afe}. {\it Top}: All stars with $\feh\geq-0.8$.  {\it Middle}: Stars with $-1.3\leq\feh<-0.8$.  {\it Bottom}: Stars with $\feh<-1.3$.}
\label{fig-anghist}
\end{figure*}

In Figure~\ref{fig-ang}, we plot each [Fe/H]-[Mg/Fe] bin in the \EzJz\ plane.  As noted in Figure~\ref{fig-accretedsims}, an orbit with high $z_{\rm max}$ and high $e$ corresponds to that with low $\jzjc$ and low $\ezec$ in this plane.  To empirically compensate for sample selection biases, as well as the uncertainty in [Mg/Fe], we directly compared the two $\alpha$-abundance regimes using their distributions in the $[\ezec,\jzjc]$ space within each metallicity bin (as described in  \S\ref{sec-bias}). These are shown in Figure~\ref{fig-anghist}. We further quantified the difference between the two regimes using a two-sample Kolmogorov-Smirnov (K-S) test, the results for which are shown in Table~\ref{tab-par}.  If the two regimes are different and we see an ``over-density" of low-$\alpha$ stars with low $\jzjc$ and low $\ezec$ as compared to the high-$\alpha$ stars, this suggests we are seeing evidence of an accreted disc component.

In the following sections, we take a closer look at our findings for each [Fe/H]-[Mg/Fe] bin.

\subsection{Searching for accreted-disc stars}

\subsubsection{$\feh\geq-0.8$}

At these metallicities, the high- and low-$\alpha$ regimes show very similar distributions in $\jzjc$, however, the high-$\alpha$ stars extend to slightly lower $\ezec$ than the low-$\alpha$ stars.  This difference in the distributions in $\ezec$ is further confirmed by the K-S test (p-value~$\sim0.005$). However, these differences do not indicate the presence of accreted disc stars. The high-$\alpha$ stars appear to be on orbits which extend slightly further out of the Galactic plane, which is similar to the distribution of the heated disc stars in the models.  On the other hand, the low-$\alpha$ stars mostly lie on highly circular orbits.  

Accreted disc stars are not expected to be on very circular orbits. Thus, these stars were most likely formed {\it in situ}, and either represent the metal-poor tail of the young disc (often referred to as the `thin disc') or were formed in the outer disc and migrated into the solar neighbourhood. We expect outer disc stars to have low $\alpha$-abundances (see \S\ref{sec:chemmod}), which has also been observed \citep[e.g.][]{bensby2011letter}.  However, constraining the role of radial migration in forming this population is beyond the scope of this work and will be investigated in a later paper.

\subsubsection{$-1.3\leq\feh<-0.8$}

In this metallicity range, the results for the low-$\alpha$ population suffer from small-number statistics.  However, we do now plausibly see accreted stars. The cold population remains, similar to the higher metallicity bin, but we now see also a population of hot stars with low $\jzjc$. Note that this metallicity bin corresponds to lower mass dwarf satellites that we do not expect to be efficiently dragged into the disc plane (see Figure \ref{fig-dga}). Thus, the fact that the `accreted stars' are not highly prograde makes sense. 

Although we plausibly see an accreted population (coming from lower mass mergers), the statistics remain poor. A K-S test between the high and low-$\alpha$ population produces p-values greater than 0.2 for both $\jzjc$ and $\ezec$. Thus, we cannot rule out that they come from similar distributions.

\subsubsection{$\feh<-1.3$}

Similar to that at higher metallicities, the low-$\alpha$ stars with $\jzjc>0$ sit well within the range of the heated disc component in the models.  The remaining stars all sit on retrograde orbits.   Further, it is clear in this metallicity regime that the high- and low-$\alpha$ populations are very similar.  We confirmed this with a two-sample K-S test, which resulted in a p-value of $\sim0.4$ in both $\jzjc$ and $\ezec$.  This suggests that they did not come from two different distributions.  This is expected since the abundance trends of the Milky Way and surviving dwarf satellite galaxies show extensive overlap at metallicities below $\sim-1.3$~dex (see Figure~\ref{fig-dga}).

\begin{table}
\caption{Two-sample K-S test for the similarity between the high- and low-$\alpha$ populations.}
\label{tab-par}
\centering
\setlength{\tabcolsep}{0.05in}
 \begin{minipage}{200mm}
  \begin{tabular}{@{}rrrrrr@{}}
  \hline
  & \multicolumn{2}{c}{$\jzjc$} & & \multicolumn{2}{c}{$\ezec$} \\
  \cline{2-3} \cline{5-6} \\
   Metallicity Bin & D & p-value & & D & p-value \\
          \hline
 $\feh\geq-0.8$ &  0.410 & $0.202$ & & 0.662 & $0.005$ \\
$-1.3\leq\feh<-0.8$ & 0.276 & 0.243 & & 0.232 & 0.518 \\
$\feh<-1.3$ & 0.148 & 0.395 & & 0.147 & 0.409 \\
\hline	
\end{tabular}
\end{minipage}
\end{table}

\subsection{A dearth of accreted disc stars}

In the lowest-metallicity bin ([Fe/H] < $-1.3$), we find evidence of accreted halo stars.  These stars show low $\alpha$-abundance; many move on retrograde orbits.  These results are similar in nature to the findings of \citet{nissen2010}, demonstrating that there is a low-$\alpha$, accreted component in the halo.  

In our intermediate metallicity bin ($-1.3$ < [Fe/H] < $-0.8$) we also plausibly see accreted stars, though the statistics are poor. In this metallicity range, our chemical template cannot distinguish massive from low-mass dwarf mergers (see \S\ref{sec:chemtemplate}). However, the kinematics of these potential accreted stars is very hot with $\jzjc \sim 0$ and no clear prograde bias (see Figure \ref{fig-ang} middle panel). This is expected if the material is accreted from low mass satellites that are not dragged towards the disc plane by dynamical friction (see \S\ref{sec:theory}). 

Our highest metallicity bin ([Fe/H] > $-0.8$) is where we hope to see evidence for massive accretion events; we find no stars with hot kinematics in this bin. Indeed, our low-$\alpha$ sample (those stars that should lie in the accreted disc star regime) are kinematically {\it colder} that the high-$\alpha$ `in-situ' sample. This could be explained by the low-$\alpha$ stars being either the low abundance tail of the thin disc, perhaps formed in-situ at an earlier time when the gas surface density was lower; or they could be migrators moving inwards from the outer disc (see \S\ref{sec:chemmod}). We will discuss these possibilities in more detail in future work. Here, we simply note that these stars are too cold to have been accreted from a massive satellite.

The apparent dearth of accreted stars has implications for the merger history of the Milky Way. Our admittedly crude analysis in \S\ref{sec:sampling} suggests that we would see accreted stars from $\sim 20^\circ$ mergers even if these comprise just 2\% of the stars in the Solar neighbourhood (see Figure \ref{fig-sampling}). The lack of an accreted disc star component thus indicates that the Milky Way has most likely experienced few if any recent interactions between the disc and massive satellite galaxies, with a correspondingly light accreted stellar and dark disc.  

Our results are in good accord with a host of other complementary studies of the Milky Way that suggest a relatively quiet merger history \citep[e.g.][]{hammer2007}: the latest measurements of the local dark matter density as compared to the value extrapolated from the rotation curve favour a nearly spherical or prolate dark matter halo at $\sim 8$\,kpc, leaving little room for a significant dark disc \citep[see][and references therein]{read2014}; the sharp fall-off in the stellar halo density profile suggests that the Milky Way experienced few late mergers \citep{deason2013}; the relationship between [Mg/Fe] and the velocity dispersions of stars in the Galactic disc also suggests few late $\simgt 1:10$ mergers \citep{minchev2014}; and finally, previous studies of the stellar populations in the Milky Way \citep[e.g.][]{wyse2001,ruchti2010}, suggest that $\simgt 1:10$ mergers are decreasingly important for the Galactic disc at later times.

It is possible -- perhaps even likely -- that the Milky Way did experience high-redshift ($\sim10$~Gyr ago) $\simgt 1:10$ mergers. The heating induced by these early mergers could heat a pre-existing thin disc \citep{gilmore2002,wyse2006,read2008,kazantzidis2008}, and/or drive disc heating via radial migration \citep{minchev2013} to form the thick disc of the Milky Way. But, it seems that our Galaxy has been rather quiet after this early violent phase, with little in the way of significant mergers since $z \sim 1$ ($\sim 8$\,Gyrs ago).

\section{Conclusions and Future Prospects}\label{sec:conclusions}

In this work, we constructed a chemo-dynamical template -- based on a combination of chemical and dynamical models, and spectroscopic observations -- from which we were able to identify stars accreted into the Milky Way disc from merging satellites. Should the satellite be massive enough and on a prograde orbit, we expect a significant fraction of its accreted stars (and dark matter) to be deposited into the Galactic disc due to dynamical friction plane dragging.

We applied our analysis template to the high-resolution spectroscopic sample from \citet{ruchti2011mpd}, which demonstrates the power of the $[\ezec,\jzjc]$ space as a discriminator between accreted stars and those that formed in situ. We found evidence only for accreted stars on retrograde or non-rotating orbits, with no compelling evidence for an accreted prograde disc component. Instead, our template uncovered a population of low-$\alpha$, metal-poor thin disc stars, consistent with having migrated from the outer disc to the solar neighbourhood. A deeper investigation is needed to confirm or rule out this scenario; we will consider this in forthcoming papers.

Our results suggest that the Milky Way had very few, if any, major mergers after the formation of its disc, since these would inevitably contribute to an accreted disc component that is not seen. This indicates the Milky Way had a relatively quiescent merger history (as is the case for $\sim 10\%$ of Milky Way mass halos; see \S\ref{sec:accstat}). We caution, however, that the high-resolution samples studied in this work suffer from small-number statistics, as well as, observational selection biases. Such impediments may restrict our ability to detect accreted disc stars, especially if they are rare. Thus, a much larger, and unbiased, sample is needed to confirm our findings.  

In a follow-up paper, we will apply the methodology outlined here to the large dataset from the Gaia-ESO Survey \citep{gilmore2012}.  This ongoing survey delivers high-quality spectral data for several thousands of stars using the FLAMES/GIRAFFE and UVES instruments at the Very Large Telescope (VLT).  With these survey data, we will have a large enough dataset to not only confirm our results for the solar neighbourhood, but to also probe for accreted disc stars beyond the solar neighbourhood.  Future large-scale spectroscopic surveys, such as 4MOST \citep{dejong2012} and WEAVE \citep{dalton2012}, in combination with high-precision astrometry from {\it Gaia}, will permit us to extend our analysis even further, to the extremities of the disc.

\section{Acknowledgements}
GRR and SF acknowledge support by grant No. 2011-5042 from the Swedish Research Council.  JIR would like to acknowledge support from SNF grant PP00P2\_128540/1.  TB is funded by grant No. 621-2009-3911 from the Swedish Research Council.  This work was supported in part by the National Science Foundation under Grant No. PHYS-1066293 and the hospitality of the Aspen Center for Physics. 

\appendix
\section{An alternative phase mixing time calculation}\label{alternativephasecalc}

In \S\ref{sec:theory}, we presented a simple back of the envelope calculation of the phase mixing time for stars in the outer halo. We assumed that stars have a spread of orbital periods set by the velocity dispersion of the progenitor. An alternative approach is to assume that stars leave the tidal radius of the progenitor at zero velocity. We show here that this gives exactly the same scaling as in equation \ref{eqn:deltaT}. 

As previously, we assume that a small substructure (a satellite) is being steadily unbound on a circular orbit of radius $d$ around a spherical host galaxy with enclosed mass $M(<d) = M$. Let us assume now, however, that stars leave this satellite with zero velocity from the tidal radius \citep[e.g.][]{2006MNRAS.366..429R}: 

\begin{equation} 
r_t \sim d \left(\frac{M_s}{3M}\right)^{1/3}
\end{equation}
where $M_s$ is the mass of the satellite enclosed within $r_t$. The phase mixing time will then be dictated by the difference in orbital period of stars leaving from either side of the satellite: 

\begin{eqnarray} 
\Delta T & = & 2\pi \sqrt{\frac{d^3}{GM}}\left[(1 + r_t/d)^{3/2} - (1 - r_t/d)^{3/2}\right] \nonumber \\ 
& \simeq & 6\pi \sqrt{\frac{d^3}{GM}} \frac{r_t}{d} \nonumber \\ 
& \simeq & 3 t_{\rm orb} \left[\frac{M_s}{3 M}\right]^{1/3}
\end{eqnarray} 
Now, using $v^2 \sim GM/d$ and $\sigma^2 \sim GM_s/r_t$, we have:

\begin{equation} 
\Delta T \simeq \frac{1.7 \sigma}{v} t_{\rm orb}
\end{equation}
Giving exactly the same scaling as equation \ref{eqn:deltaT}, albeit with slightly longer phase mixing times.

\section{Using a different abundance cut}\label{alternativemgfe}

In \S\ref{sec:results}, we separated the stars in R13 into two different $\alpha$-abundance regimes, cutting at ${\rm [Mg/Fe]}=0.3$.  This was based on the fact that for $\feh>-1.3$, stars in surviving satellite galaxies have [Mg/Fe] ratios typically less than $\sim0.3$~dex, while stars in the Milky Way typically have [Mg/Fe] ratios that are higher (see Figure~\ref{fig-dga}a).  Here, we show that if we instead make a cut at ${\rm [Mg/Fe]}=0.2$, our overall results, as shown in Figure~\ref{fig-ang}, remain the same.

\begin{figure}
\centering
\includegraphics[height=0.6\textwidth]{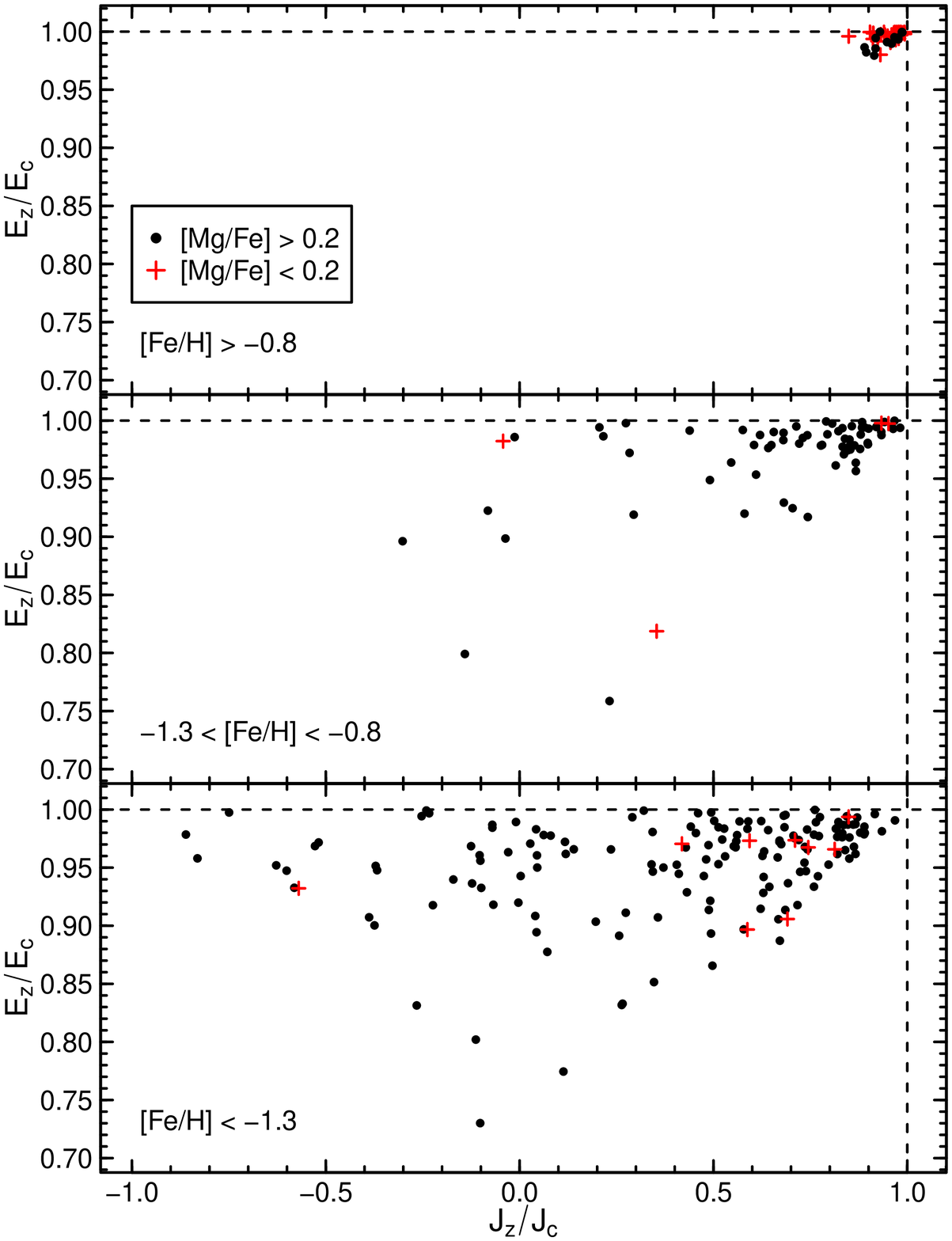}
\caption{The vertical component of the specific orbital angular momentum vs. the vertical component of the orbital energy for the R13 data set, in which we separate high- and low-$\alpha$ stars using a cut at ${\rm [Mg/Fe]}=0.2$.  These can be directly compared to the results in Figure~\ref{fig-ang}.  Those stars with ${\rm [Mg/Fe]}>0.2$ are shown as black circles while those less than $0.2$~dex are shown as red, plus symbols.  {\it Top}: All stars with $\feh\geq-0.8$.  {\it Middle}: Stars with $-1.3\leq\feh<-0.8$.  {\it Bottom}: Stars with $\feh<-1.3$.}
\label{fig-newang}
\end{figure}

In Figure~\ref{fig-newang}, we show the data plotted in the $[\ezec,\jzjc]$ plane, which can be directly compared to Figure~\ref{fig-ang}.  As expected, the cut at a lower [Mg/Fe] yields fewer low-$\alpha$ stars.  The small number of low-$\alpha$ stars makes it increasingly difficult to determine the nature of these stars.  However, our general interpretation for each metallicity bin remains the same.  We find potential accreted stars in the lowest-metallicity bins, however, we find no evidence of accretion in the highest metallicity bin.  This matches with our results when we cut at ${\rm [Mg/Fe]}=0.3$.

\bibliographystyle{mn2e}
\bibliography{accref}

\clearpage

\end{document}